\documentclass[twocolumn]{revtex4}

\usepackage{times}
\usepackage[sort&compress]{natbib}
\usepackage[dvips]{graphicx}

\begin{document}

\affiliation{$^2$}


\title{The intermediate evolution phase in case of truncated selection.}
\author{David B. Saakian$^{1,2,3}$,
Christof K.\ Biebricher $^4$,Chin-Kun Hu $^{2,5}$}
\affiliation{{}$^1$Yerevan Physics Institute, Alikhanian Brothers
St. 2, Yerevan 375036, Armenia,
\\{}$^2$Institute of Physics, Academia Sinica, Nankang, Taipei
11529, Taiwan,
\\{}$^3$ National Center for Theoretical Sciences:Physics
Division, National Taiwan University, Taipei 10617, Taiwan,
\\{}$^4$Max-Planck-Institute for Biophysical Chemistry,
D-37070 G\"ottingen, Germany
\\{}$^5$Center for Nonlinear and Complex Systems
and Department of Physics, Chung-Yuan Christian University, Chungli
320, Taiwan}
\date{\today}

\begin{abstract}
Using methods of statistical physics, we present rigorous
theoretical calculations of Eigen's quasispecies theory with the
truncated fitness landscape which dramatically limits the available
sequence space of a reproducing quasispecies. Depending on the
mutation rates, we observe three phases,  a selective one, an
intermediate one with some residual order and a completely
randomized phase. Our results are applicable for the general case of
fitness landscape.

\end{abstract}
\maketitle

\section{Introduction}
 Developing realistic evolution models poses an important
challenge for evolution research \cite{DBEH01,ei02}). After the
seminal work by Eigen\cite{ei71} and successful experiments with the
self-replication of macromolecules\cite{MPS67,Bie87}, several
theoretical studies attempted to explain this molecular evolution
phenomenon [6-18].  Realistic fitness landscapes are not smooth,
include neutral and lethal types, as observed in recent experimental
studies with RNA viruses\cite{do03,mo04,PGCM07}. In the most of
evolution articles  symmetric fitness landscapes are considered,
where the fitness is a  function of the Hamming distance from the
wild (reference) sequence. While solving evolution models, the vast
majority of results for the mean fitness have been derived using
uncontrolled approximations even for the symmetric fitness
landscapes, with too simplified sequence space \cite{wa93} and
ignoring back mutations\cite{ei89}. A simplified geometry with only
two (Hamming) classes for sequences with nonzero fitness was  used
in studies that investigate the role of lethal mutants in
evolution\cite{su06}. Furthermore, in most evolution models the
whole sequence space is assumed to be available for the evolving
genome.
 However, the sequence space a limited population can use is
severely restricted to a small part of the sequence space surrounded
by an unsurmountable moat of lethal mutations. In this paper, we
attempt to rigorously solve this case of a truncated fitness
landscape for symmetric fitness landscape.

\section{The system}
 In Eigen's theory\cite{ei71,ei89},
an information carrier reproduces with a certain rate $r_i$,
producing offspring of the parental type with the probability
$Q_{ii}$ and offspring of another (mutant) type $k$ with the
probability $Q_{ki}$,
\begin{equation}
\label{e1} dx_i/dt = \{Q_{ii}r_i - \sum_{k}r_k x_k(t)\}x_i(t) +
\sum_{k\ne i}Q_{ik}r_kx_k(t)
\end{equation}
The whole sequence space contains $M=4^L$ different sequences, where
$L$ is the genome length. The mean fitness of the system is
$R=\sum_{k}r_k x_k(t)$
 where $x_i$ are the relative frequencies of the different
genotypes ($\sum_{k=0}^{4^L-1}x_k=1$). The master type $0$ has the
maximal fitness $r_0$, and its copying fidelity is $Q\equiv Q_{00}=
q^L$, where $ q$ is the average incorporation fidelity and $L$ the
chain length. It is convenient to work with the error rate $u\equiv
L(1- q)$, leading to $Q=e^{-u}$. In this article we consider only
the case $L\to \infty$, $u$ is finite. Eigen maps each genotype
precisely into a node on the L-dimensional hypercube and has thus
the correct connectivity for each type. The minimal number of steps
leading from one position $i$ in sequence space to another one, $j$,
is the Hamming distance $d_{ji}$.

 In the evolution process the information content of the population can be
maintained only when the selection force is higher than the
dissipating one (mutation). Otherwise, above the \emph{error
threshold}, the information gets lost.

It has been shown that the system of nonlinear  differential
equations in Eq.~\ref{e1} can be transformed to an infinite system
of linear equations, connecting Eigen's model with statistical
mechanics\cite{leut87,ta92}. For a single peak fitness landscape
($r_0=A$ and $r_{i\ne 0}=1$), the following condition for conserving
the master sequence in the population holds \cite{ei71,ei89}:
\begin{equation}
\label{e2} AQ>1
\end{equation}
where $AQ=1$ is the error threshold.

At the selective phase one has \cite{dr01}
\begin{equation}
\label{e3} x_0=\frac{QA-1}{A-1}\sim 1,
\end{equation}
and   $x_i\sim 1/L^d$ for $d\ll L$, where $d$ is the Hamming
distance from the wild sequence, see Eq. (21) in \cite{sh06}. We
choose the sequences with $1\le l\le L$ from the corresponding
$l$-th  Hamming classes.
 A  scaling  by Eq.(3) exists also for the
rugged (Random Energy Model like) fitness landscapes [11]. Scaling
like the one in Eq. (3) has been applied in models of population
genetics with few alleles.
 In realistic fitness landscapes, however, the wild type is
present only in a few percents. Assuming neutrality, we can attain
such scaling: a substantial fraction of one mutation neighbors of
the wild sequence have the same high fitness.
 Neutrality increases the probability
of such mutants and suppresses $x_0$ as low as
\begin{equation}
\label{e4} x_0\sim 1/\sqrt{L}.
\end{equation}
This result could be derived easily using Eq. (6) in \cite{nc99},
for the case when there is a central neutral sequence and large
fracture of neutral sequences among the neighbor sequences of the
central sequence.

In non-selective phase one has
\begin{equation}
\label{e5} x_0\sim 1/M,
\end{equation}
where $M$ is the total number of sequences, $M=4^L$.

The error threshold phenomenon closely resembles the
ferromagnetic-paramagnetic phase transition\cite{ei89}, where the
fitness of the system corresponds to the microscopic energy of the
physics system, the mean fitness of the quasispecies to the free
energy, and the mutation rate to temperature. To identify the
different phases in statistical physics one uses the free energy and
also the order parameters.
 A phase transition occurs when, during a change
of temperature, the analytical expression of the free energy
changes. Order parameter changes also: while magnetization is
non-zero in the ferromagnetic phase, it is zero at high temperatures
and in the absence of a magnetic field.
 A phase transition in
evolution is identified by observing the mean fitness
$R=\sum_ix_ir_i$ and choosing proper order parameters, for instance
the degree of distribution around the master sequence, the surplus
production, $s=\sum_ix_i(1-2d_{0i}/L)$, where $d_{0i}$ is the
Hamming distance from wild type.

Instead of the 4-letter alphabet of genotypes, we consider only two
symbols in a genome, the spins "+" and "--", thus now $M=2^L$ [8].
Base substitutions correspond to sign changes of the spins. It is
particulary easy to analyze landscapes where the fitness values are
simple functions of the Hamming distance [8]. The N-dimensional
sequence space is then transformed into a quasi-one-dimensional,
linear chain of mutant classes $l$ where $p_l=N_lx_l$ comprises all
types with the Hamming distance number $l$ from the master and the
fitness value $J_l$. $ N_l=\frac{L!}{l!(L-l)!}$ is the number of
different genotypes in the class. The parameter $l$ can be
identified as a phenotype parameter.

There is a principal difference between quasi-one dimensional model,
derived rigorously from the initial sequence space with $2^N$
sequences and the one-dimensional one considered in [22] and other
articles.
 In contrast to other one-dimensional models
used earlier \cite{wa93} where each class contains only one type, in
our case any class $l$ is composed of $N_l$ types and thus retains
the connectivity; the Hamming distance between two sequences in the
same class can take any value from $0$ to $2l$. Moreover, when
evolution equations are formulated for class probabilities, the
effective mutation rates to the lower class, $\sim (L-l)/L$, and to
the higher class, $l/L$ are different and change with $l$
\cite{bw01},\cite{hi96}. In contrary, in the one-dimensional model
of [22] these mutation rates are l-independent.

In our quasi-one dimensional model $J_l$ can be transformed into the
$f(k)$ where $f(k)$ is an appropriate smooth function with the
maximum at $k=1$, and $f(0)=1$. The "magnetization" parameter $k$ is
defined as $k\equiv(1-2l/L)$. A correct version of 1-dimensional
evolution model has been suggested first in \cite{hi96}, the
discrete time version of parallel model with a linear fitness.

Consider now the solution of the Eigen model with symmetric fitness
landscape.
 The mean fitness $R$ for the fitness function $f$  has
been derived as follows: \cite{sh06}
\begin{equation}
\label{e6} R=\max\{f(k)Q^{(1-\sqrt{1-k^2})}\}|_{-1\le k\le 1}.
\end{equation}
$s$ can be identified from the mean fitness expression using an
equation
\begin{equation}
\label{e7}f(s)=R
\end{equation}
 as
has been derived in \cite{bw01} for the parallel model. Thus in
Eq.(6) the maximum is at some $k_0$, an order parameter of the
system quantifying the bulk spin magnetization, while the surplus
$s$ corresponds to the surface magnetization. Eq.~(6) is an exact
expression (at the infinite genome limit), while in other studies
\cite{ei89,su06} back mutations have been ignored. As shown by
Tarazona\cite{ta92}, the Eigen model is not equivalent to the simple
ferromagnetic system of spins in the lattice, but only to those
spins interacting both inside the bulk of the lattice, and on the
surface of the lattice. In this work, different phases will be
characterized by $R$, the mean fitness, by $k_0$, the bulk
magnetization, by $x_0$, the fraction of the wild type of the total
population and by $s$, the surplus. When $k_0=0$, resulting in
$s=0$, the population spreads statistically in sequence space,
indicating a non-selective phase.

We gave the mean fitness and error threshold (when $s$ in Eq. (7)
becomes 0) for the symmetric fitness landscape. The point is that
this transition has also information theoretical meaning. Eigen
actually
 found the error threshold from information theoretical
consideration of his model. Eigen's idea  (information theoretical
content of a model) resembles the investigation of information
theoretical (optimal coding) aspects of  disordered systems,
developed in statistical physics two decades later\cite{so89,sa92}.
In the Random Energy Model of spin glass \cite{de81} the phase
transition point was derived using the information theory analogy
\cite{so89,sa92}, and was found to yield results corresponding to
those derived by Eigen. The deep information theoretical meaning of
error threshold transition in evolution models (equivalent to
Shannon inequality for optimal coding) is a solid argument that
transition like the one by Eq. (2) exists for
 any (irregular, with lethal or neutral mutants) fitness landscapes.

\section{Wagner \& Krall theorem.}
{Wagner \& Krall \cite{wa93} considered a population composed of the
master and an infinite linear chain of mutants, where each type
mutates only to its next neighbor and the fitness $r_i$ decreases
monotonically. When there is no low bound of the fitness, an absence
of the error threshold transition was derived. Indeed, when in
Eq.~\ref{e6}  $f(0)= 0$, there is no error threshold transition. But
in more general symmetric fitness landscapes with a finite $f(0)$,
this ceases to be valid. The proof is as follows: the maximum of
types are located at the Hamming distance class $L/2$ or,
equivalently, at $k=0$. Consider the logarithm of the right hand
side in Eq. 6, and expand near $k=0$:
\begin{equation}
\label{e8} (1-\sqrt{1-k^2})\ln Q+\ln f(k)\approx -u\frac{k^2}{2}+\ln
(f(0))+ck^\epsilon
\end{equation}
where $c$ and $\epsilon$ are parameters describing the function
$f(k)$, and $\ln f(k)-\ln f(0)\sim ck^\epsilon$ at $k\to 0$.  When
the fitness decreases slowly and so $\epsilon < 2$, Eq.~\ref{e8} has
a maximum at $k>0$, fulfilling the condition for selection. When
$\epsilon \ge 2$, it can be demonstrated that
 there is a maximum at $ k=0$ for a sufficiently low
reproduction fidelity $Q$, therefore a sharp error threshold
transition results.
 In the too simplistic model of Wagner \& Krall, the right hand side
of Eq.~\ref{e8} lacks the quadratic term, resulting in a monotonic
function of $k$ and the absence of phase transition. In the Eigen
model the quadratic term holds, breaking the monotonic character of
 $R$ in Eq.~\ref{e8} and
invoking the error threshold.

\section{Truncated single peak fitness landscape.}
Let us consider a symmetric fitness landscape, where there is
non-zero fitness only to some Hamming
  distance from the reference
sequence.   Here we define
 the truncated
landscape as a single-peak one where all sequences beyond the
Hamming distance $d\equiv L(1-K)/2$ are lethal:
\begin{equation}
\label{e9} r_l=1 \quad 1\le l\le d, \qquad r_l=0 \quad l>d
\end{equation}
Now we have
\begin{equation}
\label{e10} M=\frac{L!}{d!(L-d)!}
\end{equation}
non-lethal sequences.

To define the mean fitness, we compare the expression
 of Eq. (6) inside the region $K<k< 1$ and at the border.

The investigation of this model is instructive, see  Figs.~1-4.
\newline
When $QA>1$ the phase is selective (phase I), $x_0$ is given by
Eq.(3), $k_0=1$, $R=QA$, see \cite{ki66}.
\newline
When
\begin{equation}
\label{e11}1>QA>Q^{1-\sqrt{1-K^2}},
\end{equation}
  a new  phase II prevails with
\begin{equation}
\label{e12}k_0=1,R=QA
\end{equation}
In the II phase $x_0$ is decreasing exponentially with L. The
expression of $x_0$ is calculated in the appendix,
\begin{eqnarray}
\label{e13}
 x_0=\exp[\int_K^1dm \frac{1}{2}\ln
\frac{\frac{\ln A}{u}+\sqrt{\frac{(\ln A}{u})^2-1+m^2}}{1+m}]
\end{eqnarray}
 At
the the transition point (to the II phase) the expression in the
exponent is becoming zero, therefore the transition is continuous.
When
\begin{equation}
\label{e14}QA<Q^{1-\sqrt{1-K^2}},
\end{equation}
  the maximum is at the border and we have
the non-selective phase III,
\begin{equation}
\label{e15}k_0=K,R=Q^{1-\sqrt{1-K^2}}
\end{equation}
 The expression for $x_0$ is defined in the appendix, Eq.(A14).
 There is some focusing around reference sequence, and $x_0$ is
 higher than $1/M$. For $K=0.9$ we obtained $x_0\approx 1/M^{2/3}$.
\newline
The transition between II and III phases is a discontinuous one,
$x_0$ decreases $M_1$ times, see Eq. (A21)
\begin{eqnarray}
\label{e16} M_1\sim  \exp[L\int_K^1dm \frac{1}{2}\ln
\frac{\sqrt{1-K^2}+\sqrt{m^2-K^2}}
     {\sqrt{1-K^2}-\sqrt{m^2-K^2}}]
\end{eqnarray}
Fig.4 illustrates  different behavior of $x_0$ in three phases.

While in the ordinary Eigen model (without truncation of fitness)
there is a sharp phase transition with the jump of the $x_0$
behavior from the Eq.(3) to the $1/M$, in truncated case these sharp
transition is moved to the transition point between II and III
phases. Now $x_0$ is continuous at transition point between I and II
phases. For the Summers-Litwin case with $K=1$ we have $M_1\to 1$,
therefore the transition disappears as has been obtained in
\cite{su06}, and for the $K=0$ case we get the result of Eigen model
$M_1=1/2^N$. Our formulas are derived for the case $K\ge 0$.

Is phase II a selective one in the ordinary meaning?
 Clearly, its mean fitness is higher than
in a typical non-selective phase like phase III. This point was
clarified by calculating the surplus, replacing the step-like
fitness function near the borderline ($k=K$) with a smooth function
$f(k)$ which changes its value from 1 to 0 near $K$. In both phases,
II and III, with $QA>Q^{1-\sqrt{1-K^2}}$, the majority of the
population is near the borderline, both on the viable and the lethal
side. Therefore, while there is a kind of phase transition with some
population rearrangement, phase II is identified as an intermediate
one,  with $x_0\ll 1$, as in the non-selective phase. Summers and
Litwin \cite{su06} first realized that $k_0\to 1$ in a truncated
fitness landscape and tried to analyze the phenomenon.
Unfortunately, they used too simplistic a model where all mutants
except the nearest neighbors of the master type were lethal.
 Fig.~2  compares the relative concentration of the
master at various superiorities in the single-peak fitness landscape
described in \cite{ei71,ei89}, with the truncated  fitness landscape
 with $d=8$, and the case considered by Summers and
Litwin. Note the strong dependence of the master concentration on
the master superiority in the more realistic landscape, leading to
an error threshold, in contrast to the case in \cite{su06}. The
population profiles of the truncated fitness landscape at different
mutation rates are shown in Fig.~3, showing the transition from a
master-dominated population via a wide-spread mutant distribution to
a non-selective case.

In the Table1 we give the results of numeric for different phases.
Mean fitness is well confirmed numerically, while the accuracy of
numerics is poor to get correct values of $x_0$ in the second phase,
$2.37<A<2.77$.
\begin{figure}
\centerline{\includegraphics[width=0.6\columnwidth]{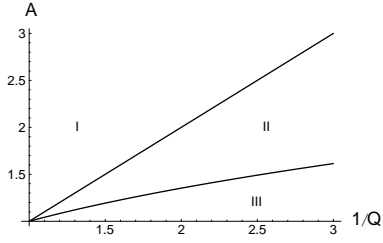}}
\caption{ Phase structure for the model with the overlap parameter
for truncation point $K=0.9$, and mutation rate  $u=1$. I: selective
phase with $QA>1$, $A$ is the fitness at the peak, $Q=e^{-u}$ is the
errorless copying probability of the genome. II: Intermediate phase
with $QA<1$ and $QA>Q^{1-\sqrt{1-K^2}}$. III: non-selective phase
with $QA<Q^{1-\sqrt{1-K^2}}$.} \label{fig2}
\end{figure}
\begin{figure}
\centerline{\includegraphics[width=0.6\columnwidth]{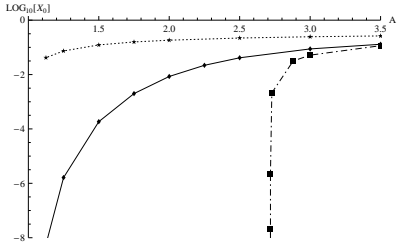}}
\caption{  Semilogarithmic plot of $x_0$ versus $A$ for a single
peak fitness landscape truncated at $d>8$ with $u=1,L=3000$. The
upper curve corresponds to the Summers-Litwin model, the middle
curve to the truncated fitness landscape as in Fig.~3. The square
boxes correspond to a single peak fitness model. For $\ln(A)<1$
Eigen model with single peak fitness has $x_0=0$.} \label{fig2}
\end{figure}
\begin{figure}
\centerline{\includegraphics[width=0.6\columnwidth]{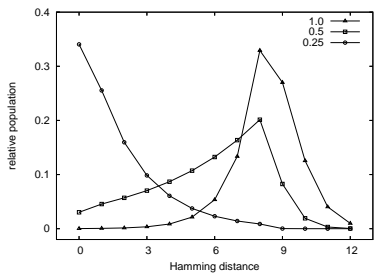}}
\caption{Truncated Eigen model with $d=8$, $A=3,u=1, L=3000$. The
distribution of probabilities of Hamming classes $p_l$ at different
mutation rates ($u$ values). At high accuracy (circles) the
population groups around the master type, at intermediate accuracy
the population peaks at the truncation border, at low accuracy the
master type has practically disappeared and the majority of the
progeny is lethal. } \label{fig2}
\end{figure}
\begin{figure}
\centerline{\includegraphics[width=0.6\columnwidth]{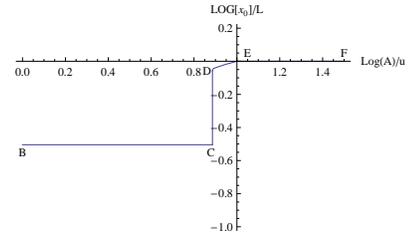}}
\caption{The graphics of $\ln x_0/L$ for the truncated Eigen model
at infinite $L$ with $K=1/2,u=1$. The I phase (curve BC) is at
$1<\ln A/u$, the second (curve DE) at $1<0.88<\ln A/u<1$ and the
third (curve EF)at $0<\ln A/u<0.88$. The curve CD corresponds to the
jump at the border between II and III phases. In the first phase
$\ln x_0/L\sim O(1/L)$.} \label{fig2}
\end{figure}

\begin{center}
\begin{table}[t]
\begin{tabular}{|c|c|c|c|c|c|c|c|c|c|}
\hline Phase        &III  &III  &II   &II   &II     &II             &I      \\
\hline A             &2.2  &2.3  &2.4  &2.5  &2.6    &2.7            &2.8     \\
\hline R             &0.859&0.859&0.883&0.919&0.957  &0.993          &1.030   \\
\hline R th.      &0.875&0.875&0.882&0.919           &0.956     &0.992          &1.029   \\
\hline $x_0$   &  $4*10^{-117}$   &   $7*10^{-54}$  & $2*10^{-24}$    & $4.7*10^{-16}$  &$10^{-9}$&$4.4* 10^{-4}$&0.0167\\
\hline $x_0$ th.  &     &     &     &     &       &               &0.0164   \\
\hline
\end{tabular}\caption{ Eigen model with truncated single peak fitness landscape for $L=1000,u=1,K=1/2$.
 The transition between I and second phases is at $A\approx 2.778$,
between II and III phases at $A\approx 2.377$.}
\end{table}
\end{center}
How the phases can be identified? The fact is that the parameters
$k_0$ and $s$ have different meanings in a statistical physics
approach. This subject has been well analyzed in a series of
articles by E. Baake and her co-authors. $s$ and $k_0$ have been
identified with the transverse and longitudinal magnetizations of
spins in the corresponding quantum model.
We just link $s$ with the mean characteristic of the phenotype, and
$k_0$- with the repertoire of genotypes.
 The
consensus sequence should  be determined experimentally not only via
distribution $x_i$, but also via distribution
$(x_i)^2/\sum_{j=0}^M(x_j)^2$. Consider
\begin{equation}
\label{e17}\hat k_l=\frac{N_l(x_i)^2}{\sum_j(x_j)^2},
\end{equation}
where $x_i$ belongs to the phenotype class $l$ with $N_l$ different
genotypes. Having such data, one can simply identify the phase
structure.

In the appendix we solve the truncated fitness models for the
general monotonic smooth function $f(x)$. The numerics confirms well
our analytical results for the new phase.

\section{Discussion}
We rigorously solved (at the infinite genome length limit) Eigen's
model for the truncated selection using the method of \cite{sh06} as
well as methods of statistical mechanics, including the analogy of
the error threshold to the ferromagnetic-para\ -magnetic transition.
This analogy is a complicated critical phenomenon, presented by
Leuth\"ausser and Tarazona \cite{leut87,ta92} and well analyzed by
E. Baake and co-authors \cite{ba97,bw01}. Instead of using only one
order parameter to identify the phase of the model, magnetization,
it is necessary to take into account several order parameters
describing the order of spins in the bulk lattice and at the
surface. Recently the existence of an error threshold has been
questioned \cite{su06} in the case of truncated selection.
 In this model the
available sequence space has been shrunk to an extremely small size.
Fig.~2 illustrates that the unrealistic assumption in \cite{su06}
changed the relative concentration of the master type by more than 3
orders of magnitude. Nevertheless, this work was certainly useful
for clarifying the concept of quasispecies: the authors first
realized the intriguing features of a truncated selection landscape.
We found a new evolution (intermediate) phase, when there is no
successful selection via  phenotype trait (the majority does not
share the trait), while there is some grouping of population at
genotype level. The intermediate evolution phase differs from the
non-selective phase, the frequency of wild type being
$\frac{1}{\sqrt{M}}$ or higher in the intermediate phase compared
with $\sim\frac{1}{M}$ in the non-selective phase of Eigen model. We
proposed a parameter to measure the hidden grouping of population in
a genotype level, Eq. (17). Such hidden ordering could be important
in case of changing environment: it is possible to force the whole
population to extinction changing the fitness of a small fraction
(much smaller than $ 1/L$, but much higher than  $1/M$, M is the
number of different genotypes) of viruses in the population.  We
recommend virologists to measure the consensus sequence not only
using the the probabilities $x_i$, but also $x_i^2$. The evolution
picture of the virus population is robust when two versions of
consensus sequence are close to each other.
  In experiments \cite{do03}
has been observed an evolution picture, qualitatively similar to the
intermediate evolution phase.

 How the error threshold transition is connected to the virus
 extinction in virus experiments, is another story.
 Several mechanisms are possible: an error catastrophe, as well
 as a critical mean fitness in order to
 maintain viral growth \cite{wi07}. In this work, we observed
 the new phase with single peak and symmetric landscapes,
 but this phase exists probably for
any (including irregular) fitness landscape with a lethal wall in
sequence space.

\section{Acknowledgments}
 D.B.\ Saakian thanks the Volkswagenstiftung grant "Quantum
Thermodynamics'' and
 the
 National Center for Theoretical Sciences
in Taiwan and Academia Sinica (Taiwan) under Grant No. AS-95-TP-A07
for the financial support. We thank E. Domingo and M.W.\ Deem
discussions.

\renewcommand{\theequation}{A.\arabic{equation}}
\setcounter{equation}{0}
\section*{Appendix A. Application of HJE for truncated symmetric landscape.}

Let us apply the Hamilton-Jacobi equation (HJE) method
\cite{sa07,sa08} to Eigen model with truncated symmetric fitness
landscape. Consider a piecewise smooth, monotonic fitness function
$f(m)$,
\begin{eqnarray}
\label{e17}f(m)=f_0(m),m>K,\nonumber\\
f(m)=0, m<K
\end{eqnarray}\begin{center}
\begin{table}[t]
\begin{tabular}{|c|c|c|c|c|c|c|c|c|c|}
\hline Phase       &III&III& II  & II        &II      &I        &I                        &I  \\
\hline c        &0.5   &  1   &1.2       &1.3        &1.5       &1.9       &   2.1                  &3  \\
\hline ln(R)    &-0.090&-0.016&0.0163&    0.035   &0.084&  0.212       &0.287                  &0.665  \\
\hline ln(R) th &-0.071&-0.009&0.0160      & 0.034   &0.083 &  0.213          & 0.288                 &0.666   \\
\hline s        &0.507 &    0.507&            0.507&0.508      &0.510  &0.525      &0.541                  &0.663  \\
\hline s th.    &0.5   & 0.5 &0.5      &0.5         &0.500& 0.5         &0.523                  &0.666   \\
\hline $-\ln(x_0)$     & 223 &199           & 185       & 177    &  160 & 131      &120                     &84\\
\hline
\end{tabular}\caption{ Eigen model with truncated fitness, $K=1/2,L=500,u=1,f(x)=\exp(cm^2/2)$.
 The transition between I and second phases is at $c=2$,
between II and III phases at $c=1/\sqrt{1-K^2}\approx 1.1547$. In
the I phase we have: $s=1/c,\ln(R)/u=c(1-1/c)^2/2$. In the II phase:
$s=K,\ln(R)/u=c(1-1/c)^2/2$. In the III phase:
$s=K,\ln(R)/u=cK^2/2+\sqrt{1-K^2}-1$. }
\end{table}
\end{center}
\begin{center}
\begin{table}[t]
\begin{tabular}{|c|c|c|c|c|c|c|c|c|c|c|}
\hline Phase       &III        &III         & III   & II     & II&  II     &I          &I                          \\
\hline c        &0.5            &  1         &2.2   & 2.4    &2.5 & 2.6   &3         &4                      \\
\hline ln(R)    &-0.135         &-0.016      &-0.022& 0.0202  & 0.0433 &  0.0676     &0.172    &0.459                  \\
\hline ln(R) th &-0.11          & -0.092     &-0.042& 0.0201      &0.0435&  0.0678   &0.171     & 0.460                        \\
\hline s        &0.507          &0.507       &0.510 & 0.514  &0.518 & 0.523      &0.563      &0.698                    \\
\hline s th.    &     0.5       & 0.5        &0.5   & 0.5   &0.5      & 0.5      &0.556      & 0.701              \\
\hline $x_0$    &230            &216         &142   & 128   &122     & 117   &98         &70\\
\hline
\end{tabular}\caption{ Eigen model with truncated fitness, $K=1/2,L=500,u=1,f(x)=\exp(cm^3)$.
 The transition point is at $c\approx 2.31,m_0=0.866$ in the model without truncation.
 The transition point between II,III is at $c\approx 2.79$.}
\end{table}
\end{center}
Here $f_0(m)$ is a monotonic analytical function.

 We denote by $k_0$ the
maximum point in Eq.(6). As $s$ is defined by Eq. (7), for the
monotonic fitness function we obtain
\begin{eqnarray}
\label{e18}s\le k_0,
\end{eqnarray}
and there is only one solution of Eq. (7).

Using an ansatz
\begin{eqnarray}
\label{e19}p_l=\exp[ LU_0(m,t)],
\end{eqnarray}
$m=1-2l/L$, in \cite{sa07} has been derived the following equation
\begin{eqnarray}
\label{e20}
L\frac{\partial U_0(m,t)}{\partial t}=f_0(m)e^{-u}\times \nonumber\\
\exp\{u[\cosh(2\frac{\partial U_0(m,t)}{\partial
m})+m\sinh(2\frac{\partial U_0(m,t)}{\partial m})]\}
\end{eqnarray}
The HJE approach works for any   piecewise smooth fitness functions.
Eq. (A1) is such a case.
 To solve our truncated case we should just use different analytical
solutions for the regions $-1<m<K$ and $K< m<1$.

Assuming an asymptotic $U_0(m,t)=\frac{R}{L}t+U(m)$, we derived
\cite{sa07}
\begin{eqnarray}
\label{e21}
R=f_0(m)e^{-u}\times \nonumber\\
\exp\{u[\cosh(2\frac{dU(m)}{dk})+m\sinh(2\frac{dU(m)}{dm})]\},
\end{eqnarray}
where $R$ is derived by Eq.(6). The surplus $s$ is defined as the
value of $m$ where $U(m)$ has a maximum. When $s$ is inside the
region $[K,1]$, $U'(s)=0$.  At extremum points with  $U'(m)=0$
Eq.(A5) gives $f(s)=R$. As for a monotonic fitness function there is
a single solution for Eq. (6), $U(x)$ has a single maximum point in
this case, therefore it is a concave function and we take $U(s)=0$.

We use Eq.(A5) to define $p_l$ with an accuracy $O(1)$ for $\ln
p_l$, calculating $U(m)=U(s)+\int_s^mU'(m)dm$ for the corresponding
$m=1-2l/L$. Moreover, it is possible to calculate $\ln p_l$ with a
higher accuracy $O(1/L)$. In [18] we gave explicit formulas for the
case of parallel model. It is possible to construct similar results
for the Eigen model as well.

 We have two branches of solutions for Eq. (A5):
\begin{eqnarray}
\label{e22}
U'(m)=\frac{1}{2}\ln\frac{q\pm\sqrt{q^2-1+m^2}}{1+m},\nonumber\\
q=\frac{1}{u}\ln \frac{R}{f(m)}+1
\end{eqnarray}
It is a principal point the choice of different solutions. We choose
the proper branch assuming:
\begin{itemize}
\item U(m) is continuous function,
\item U'(m) is continuous function,
\item U(m) is a concave function for the monotonic fitness function $f(m)$.
\end{itemize}
 The transition between two branches
($\pm $ solutions in Eq.(A5)) is only at the point where
$q^2-1+m^2=0$ or
\begin{eqnarray}
\label{e23} R=f(m)e^{-u+u\sqrt{1-m^2}}
\end{eqnarray}
According to Eq.(6), $R$ is the maximum of the right hand side. Thus
we should choose only the branch with $"-"$ sign when $k_0$ is at
the border, $k_0=K$. When k is inside the interval $[K,1]$, then we
choose the $"-"$solution  for the interval $[k_0,1]$ and $"+"$
solution  in the interval $[K,k_0]$.

For the $V(m)=\ln (\sqrt{N_lp_l})/L$ we have another equation [18],
\begin{eqnarray}
\label{e24} R=f_0(m)e^{-u}
\exp\{u[\cosh(2\frac{dV(m)}{dm})\sqrt{1-m^2}]\}
\end{eqnarray}
The minimum of the right hand side via $V'$ just gives the
$f(m)e^{-u+u\sqrt{1-m^2}}$. Thus at the maximum point $m=k_0$ of
function $V(m)$ we have
 $V'(k_0)=0$. In this article we consider the case when Eq.(6) has a single
solution $k=k_0$. Then $V(m)$ is a concave function.

Solution of equations (A5),(A7) are simply related,[18],
\begin{eqnarray}
\label{e25} V(m)=U(m)+\frac{(1+m)\ln
\frac{1+m}{2}}{4}+\frac{(1-m)\ln \frac{1-m}{2}}{4}
\end{eqnarray}

Consider now different phases of our model. The selective one with
$K<k_0<1,K<s<1$; the non-selective one with $k_0=K,s=K$, and
intermediate one with $K<k_0<1,s=K$.

{\bf Selective phase}.

Now $R$ is given by Eq. (6) with a $K<k_0<1$. We used $"-"$ solution
of Eq.(A6) for $k_0<m<1$ and the $"+"$ solution for $K<m<k_0$. The
maximum points of both functions $U(m)$ and $V(m)$ are inside the
interval $[K,1]$. We have $U'(s)=0$ and $V'(k_0)=0$. The formulas
for the steady state distributions are the same as in \cite{sa07}.
We have a mean fitness
\begin{eqnarray}
\label{e26} R=f(k_0)e^{-u+u\sqrt{1-k_0^2}}
\end{eqnarray}
For the $p_l,m=L(1-2l/L),k_0<m\le 1$ we have an expression
\begin{eqnarray}
\label{e27} p_l=\exp[L\int_{k_0}^mdm \frac{1}{2}\ln
\frac{q-\sqrt{q^2-1+m^2}}{1+m}]\nonumber\\+ L\int_s^{k_0}dm
\frac{1}{2}\ln \frac{q+\sqrt{q^2-1+m^2}}{1+m}],\nonumber\\
q=\frac{1}{u}\ln\frac{R}{f(m)}+1
\end{eqnarray}
For $m<k_0$ we have
\begin{eqnarray}
\label{e28} p_l=\exp[L\int_{s}^{m}dm \frac{1}{2}\ln
\frac{q+\sqrt{q^2-1+m^2}}{1+m}],\nonumber\\
q=\frac{1}{u}\ln\frac{R}{f(m)}+1
\end{eqnarray}

{\bf Nonselective phase}.

Now the maximum of Eq.(6) is at the border $k_0=K$, and we have
\begin{eqnarray}
\label{e29} R=f(K)e^{-u+u\sqrt{1-K^2}}
\end{eqnarray}
 We use $"-"$ solution of Eq.(A6) for the whole interval $K<m<1$.
For the $p_l,m=L(1-2l/L)$ we have an expression
\begin{eqnarray}
\label{e30} p_l=\exp[L\int_K^mdm \frac{1}{2}\ln
\frac{q-\sqrt{q^2-1+m^2}}{1+m}],\nonumber\\
q=\frac{1}{u}\ln\frac{R}{f(m)}+1
\end{eqnarray}
and the maximum is for $p_d$ with $m\equiv \frac{1-2d}{L}=K$.

 For the single peak
fitness case ($f(1)=A$ and $f(m)=1$ for $m<1$), we have
$q=\sqrt{1-K^2}$ and
\begin{eqnarray}
\label{e31} x_0\equiv p_0\sim \exp[\frac{L}{2}\int_K^1dm \ln
\frac{\sqrt{1-K^2}-\sqrt{m^2-K^2}}{1+m}]
\end{eqnarray}
For the $1/M$ we have an expression
\begin{eqnarray}
\label{e32} \frac{1}{M}\sim \exp[\frac{L}{2}\int_K^1dm  \ln
\frac{1-m}{1+m}]
\end{eqnarray}

 {\bf Intermediate phase}.
Now mean fitness is given by Eq.(6) with some $K<k_0<1$ and $s=K$.
We used $"-"$ solution of Eq.(A6) for  $k_0<m<1$ and $"+"$ solution
for $K<m<k_0$.
 When $m>k_0$, we have
 \begin{eqnarray}
\label{e33}
 p_l=\exp[L\int_K^{k_0}dm \frac{1}{2}\ln
\frac{q+\sqrt{q^2-1+m^2}}{1+m}
+\nonumber\\
L\int_{k_0}^mdm \frac{1}{2}\ln \frac{q-\sqrt{q^2-1+m^2}}{1+m}]
\end{eqnarray}
 In case of $K<m<k_0$ we have
\begin{eqnarray}
\label{e34}
 p_l=\exp[L\int_K^mdm \frac{1}{2}\ln
\frac{q+\sqrt{q^2-1+m^2}}{1+m}]
\end{eqnarray}
We took $U(K)=0$, as the maximum of population is at the border with
$m=K$.

For the SP case we have $q=\frac{\ln A}{u}$, and
\begin{eqnarray}
\label{e35}
 x_0=\exp[L\int_K^1dm \frac{1}{2}\ln
\frac{\frac{\ln A}{u}+\sqrt{\frac{(\ln A}{u})^2-1+m^2}}{1+m}]
\end{eqnarray}
Above the transition point $\ln A=\sqrt{1-K^2}$ Eq.(19) gives
\begin{eqnarray}
\label{e36}
 x_0=\exp[L\int_K^1dm \frac{1}{2}\ln
\frac{\sqrt{1-K^2}+\sqrt{m^2-K^2}}{1+m}]
\end{eqnarray}
We see that at the transition point there is a jump, $x_0$ decreases
$M_1$ times,
\begin{eqnarray}
\label{e37} M_1\sim  \exp[L\int_K^1dm \frac{1}{2}\ln
\frac{\sqrt{1-K^2}+\sqrt{m^2-K^2}}{\sqrt{1-K^2}-\sqrt{m^2-K^2}}]
\end{eqnarray}
 For $K=0.5$ Eq.(A20) gives $\ln(x_0)/L\approx-0.057$, while  $\ln(M)/L\approx
 -0.56$.
Thus above the transition point to the third phase
\begin{eqnarray}
\label{e38} Log(x_0)\sim\frac{1}{M^{0.1}}
\end{eqnarray}
Consider the case $K=0.99$. Now we have $\ln(x_0)/L\approx-0.01$ and
$\ln(M)/L\approx -0.031$. Thus
\begin{eqnarray}
\label{e39} \ln(x_0)\sim\frac{1}{M^{0.32}}
\end{eqnarray}
Consider evolution model with general fitness function $f(x)$.
Assume that without truncation the error threshold transition is a
discontinuous one, and there is a jump from non-zero $k_0>$ in
selective phase to $k=0$ solution in Eq.(6) for non-selective phase.
Let us introduce the truncation. Choosing $K<k_0$, we have three
phases, see Table 3,  and  $x_0$ decreases $M_2$ times at the
transition point between II and III phases,
\begin{eqnarray}
\label{e40} M_2\sim  \exp[L\int_{k_0}^1dm \frac{1}{2}\ln
\frac{\sqrt{1-K^2}+\sqrt{m^2-K^2}}{\sqrt{1-K^2}-\sqrt{m^2-K^2}}]
\end{eqnarray}
If in the original (without truncation) model the error threshold
transition is a continuous one with $k_0=0$, after truncation we
have different expressions for $x_0$ in the II ($k_0>K,s=K$) and III
($k_0=K,s=K$) phases while continuous transitions $I\to II $ and
$II\to III$, see Table 2. We have a similar behavior for the phase
transitions in case of originally (without truncation) discontinuous
error threshold transition, if the truncation parameter $K$ is
chosen too large,$K>k_0$.

Let us derive an important constraint for the population of the
class at the Hamming distance $n=L(1-k_0)/2$. For the corresponding
$V$ we have
\begin{eqnarray}
\label{e41} V(k_0)\equiv \frac{1}{L}\ln \frac{p_n}{\sqrt{N_n}}
\end{eqnarray}
 As $p_d=1$ (the majority of
population is at the border with the overlap parameter
$K=(1-2d)/L$), we have
\begin{eqnarray}
\label{e42} V(K)\equiv \frac{1}{L}\ln \frac{1}{\sqrt{M}},
\end{eqnarray}
as $M=N_K$. We proved before that $V(m)$ has a single maximum (in
our case with a  single solution for the maximum point $k$ in Eq.
(6)). Thus
\begin{eqnarray}
\label{e43} V(K)>V(k_0)
\end{eqnarray}
which gives
\begin{eqnarray}
\label{e44} p_n> \frac{\sqrt{N_n}}{\sqrt{M}}
\end{eqnarray}
The last inequality supports the choice of order parameter in Eq.
(17).

For the single peak case $n=0$, and we get from Eq.(A28)
\begin{eqnarray}
\label{e45} x_0> \frac{1}{\sqrt{M}}
\end{eqnarray}
Eqs.(A22,A23) give even higher values for $x_0$.


\begin{thebibliography}{99}
\bibitem{DBEH01}E. Domingo, C. K. Biebricher, M. Eigen, and J.J.
Holland, \textit{Quasispecies and RNA Virus evolution: Principles
and Consequences.} (Landes Bioscience, Austin, TX, 2001).
\bibitem{ei02}M. Eigen,
Proc. Natl. Acad. Sci. USA \textbf{99}, 13374 (2002).
\bibitem{ei71}
M. Eigen, Naturwissenschaften \textbf{58}, 465 (1971).
\bibitem{Bie87}
C. K. Biebricher, Cold Spring Harbor Symp. Quant. Biol. \textbf{52},
299 (1987)
\bibitem{MPS67}
D.R. Mills, R. L. Peterson, and S. Spiegelman, Proc. Natl. Acad.
Sci. USA \textbf{58}, 217 (1967).
\bibitem{ss82}
J. Swetina, and P. Schuster, Biophys. Chem. \textbf{16}, 329 (1982).
\bibitem{leut87} I. Leuth\"ausser,
J. Stat. Phys. 48, 343 (1987)
\bibitem{ss} P. Schuster, and J. Swetina,
Bull. Math. Biol. \textbf{50}, 635 (1988).
\bibitem{ta92} P. Tarazona,
Phys. Rev. A \textbf{45}, 6038 (1992).
\bibitem{ei89}
M. Eigen, J. S. McCaskill, and P. Schuster, Adv. Chem. Phys.
\textbf{75}, 149 (1989).
\bibitem{pe97}
S. Franz, and L. Peliti, J. Phys. A \textbf{30}, 4481 (1997).
\bibitem{ba97}
E. Baake, M. Baake, and H. Wagner, Phys. Rev. Lett. \textbf{78}, 559
(1997).
\bibitem{bw01} E. Baake, and H. Wagner, Genet. Res. {\bf 78},
93 (2001)
\bibitem{ba02}
J. Hermisson, O. Redner, H. Wagner, and E. Baake,  Theor. Pop. Biol.
{\bf 62}, 9 (2002)

\bibitem{sh06}
D. B. Saakian, and C.-K. Hu, Proc. Natl. Acad. Sci. USA
\textbf{103}, 4935 (2006).
\bibitem{sh06a}
D. B. Saakian,  E. Munoz, C.-K. Hu, and M. W. Deem, Phys. Rev. E
\textbf{73}, 041913 (2006).
\bibitem{de07}J.-M. Park and M. W. Deem, PRL, {\bf
98}, 058101(2007).
\bibitem{sa07} D.B. Saakian, Journal of statistical physics, {\bf 128},781(2007).
\bibitem{sa08}D. B. Saakian, O. Rozanova, and Andrei Akmetzhanov, Phys.Rev. E {\bf 78}, 041908 (2008).

\bibitem{mo04}
R. Sanjuan, A. Moya, and S. F. Elena, Proc. Natl. Acad. Sci. USA
\textbf{101}, 8396 (2004).
\bibitem{do03}
E. L\'azaro, C. Escarmis, J. Perez-Mercader, S. C. Manrubia, and E.
Domingo, Proc. Natl. Acad. Sci. USA \textbf{100}, 10830 (2003).
\bibitem{PGCM07}
M. Pariera, M., G. Fernandez, B. Clotet, and M. A. Martinez, Mol.
Biol. Evol. \textbf{24}, 382 (2007).
\bibitem{wa93}
G. P. Wagner, and P. Krall, J. Math. Biol.\textbf{32}, 33
(1993).\bibitem{su06} J. Summers, and M. Litwin, J. Virol.
\textbf{80}, 20 (2006).

\bibitem{dr01} B. Drossel,
Biological evolution and statistical physics
 Advances in Physics {\bf 50}, 209 (2001).
\bibitem{nc99}E.V. Nimwegen,   J.P.Crutchfield, and M. Huynen
     {\it Proc. Natl. Acad. Sci. USA} {\bf 96}, 9716 (1999).
\bibitem{hi96} H. Woodcock and P.G. Higgs, J. Theor. Biol. 179:61-73,
(1996).

\bibitem{so89}
N. Sourlas, Nature {\bf 339}, 693 (1989).
\bibitem{sa92}
D. B. Saakian,
 JETP Lett.{\bf 55},2(1992).
\bibitem{de81}
B. Derrida, Phys. Rev. B \textbf{24}, 2613 (1981).
\bibitem{ki66}Kimura and Maruyama. Genetics {\bf 54},1303(1966)
\bibitem{wi07}
J. J. Bull, R. Sanjuan, and C. O. Wilke, J. Virol. \textbf{81}, 2930
(2007).
\end{thebibliography}
\end{document}